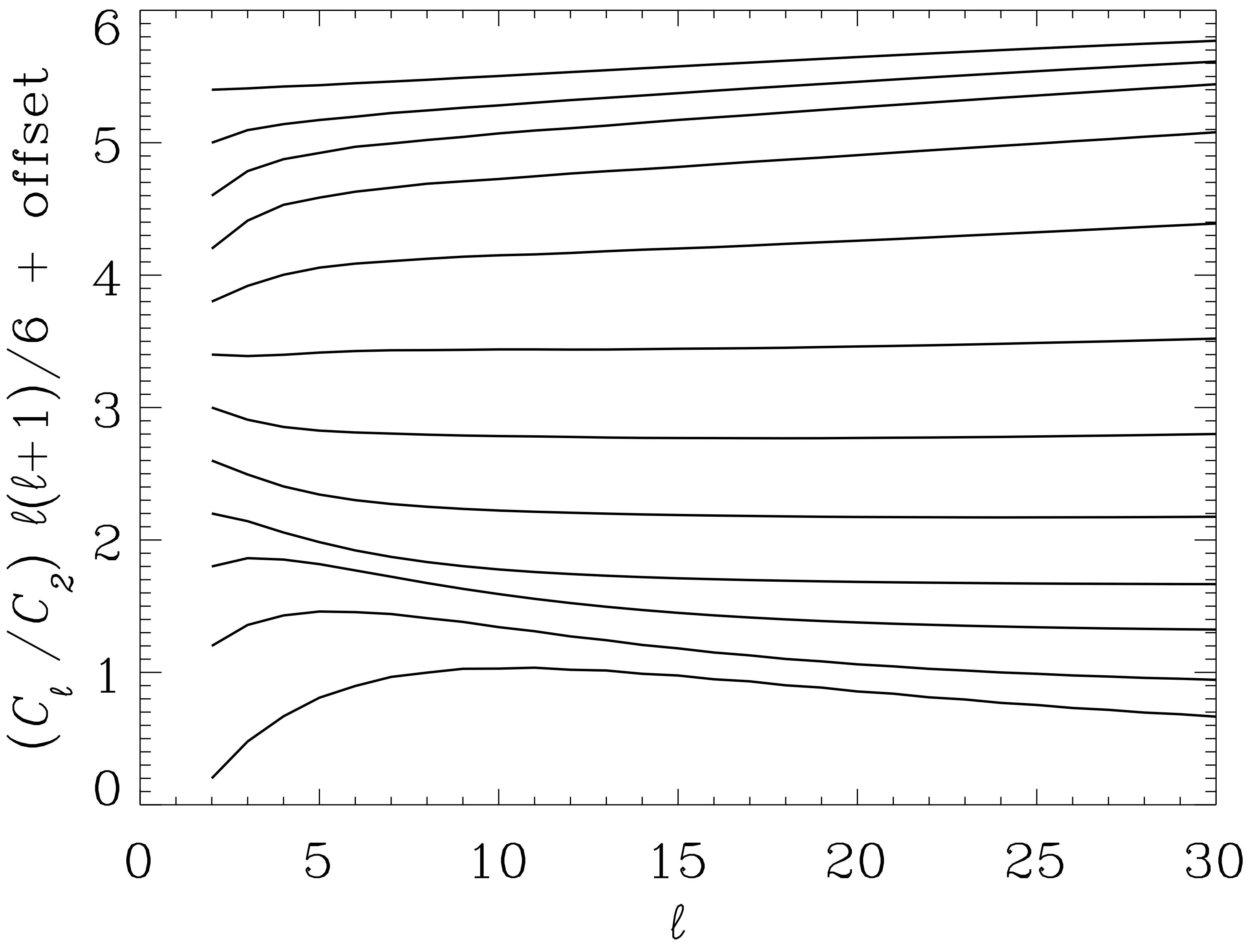

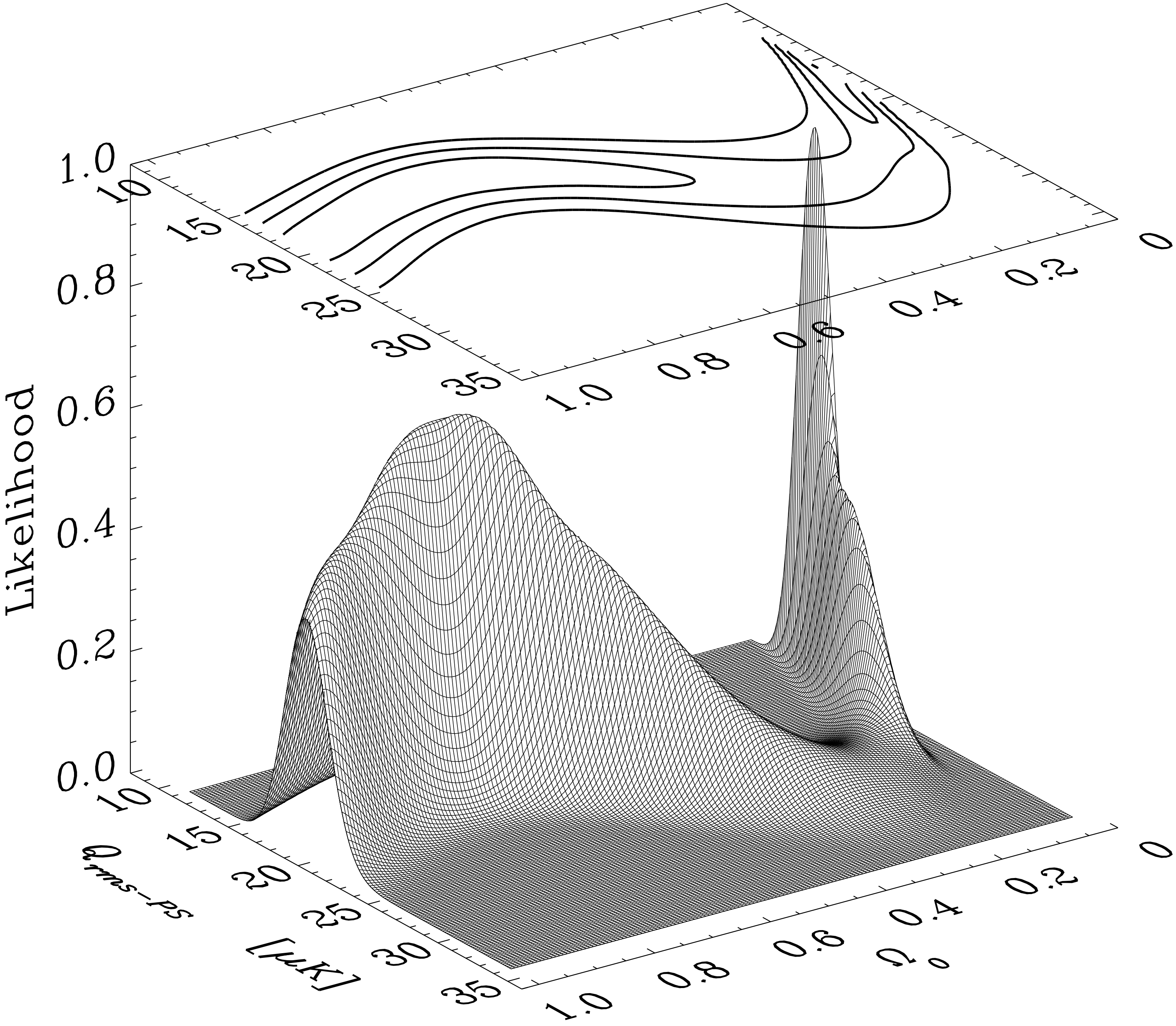

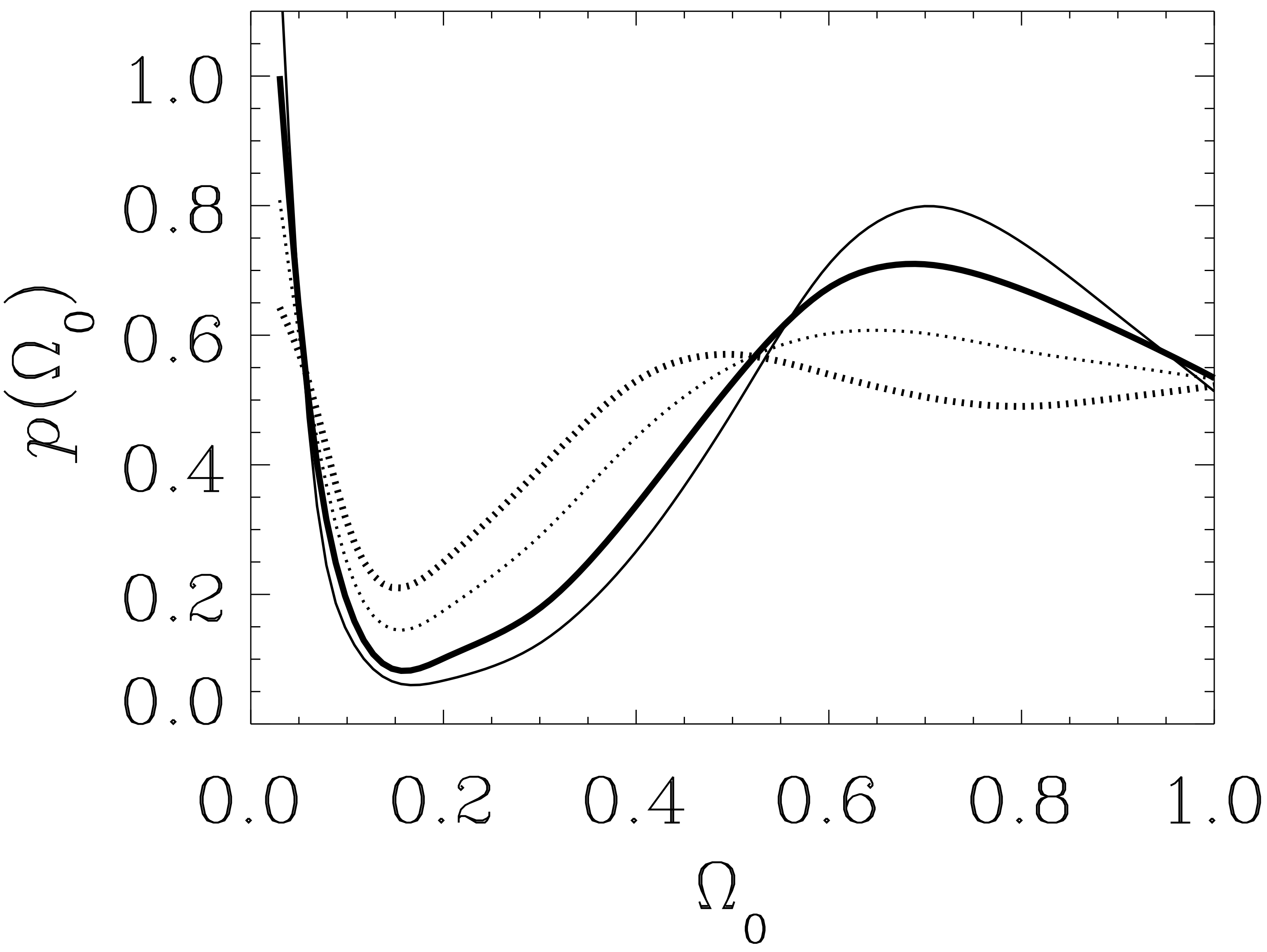

COBE-DMR-NORMALISED OPEN INFLATION, CDM COSMOGONY

Krzysztof M. Górski[1,2,6], Bharat Ratra[3], Naoshi Sugiyama[4,5], and Anthony J. Banday[1]

## ABSTRACT

Fourier analysis on a cut sky of the two-year COBE DMR 53 and 90 GHz sky maps (in galactic coordinates) is used to determine the normalisation of an open inflation model based on the cold dark matter scenario. The normalised model is compared to measures of large-scale structure in the universe. Although the DMR data alone does not provide sufficient discriminative power to prefer a particular value of the mass density parameter, the open model appears to be reasonably consistent with observations when $\Omega_0 \sim 0.3 - 0.4$ and merits further study.

[1] Universities Space Research Association, NASA/GSFC, Laboratory for Astronomy and Solar Physics, Code 685, Greenbelt, MD 20771

[2] Warsaw University Observatory, Aleje Ujazdowskie 4, 00-478 Warszawa, Poland

[3] Joseph Henry Laboratories, Princeton University, Princeton, NJ 08544

[4] Department of Astronomy, University of California, Berkeley, CA 94720

[5] Department of Physics, Faculty of Science, University of Tokyo, Tokyo 113, Japan

[6] email: gorski@stars.gsfc.nasa.gov

## 1. INTRODUCTION

Observational evidence (summarized in Ratra & Peebles 1994b, hereafter RPb) motivates the consideration of a cosmogony with a clustered mass density parameter, $\Omega_0$, significantly less than unity, but possibly somewhat larger than the baryon mass density parameter, $\Omega_B$, required for standard nucleosynthesis (e.g., Reeves 1994). In this *Letter*, we focus on a low-density cold dark matter (CDM) model with open spatial sections and zero cosmological constant $\Lambda$, with an early inflation epoch (Ratra & Peebles 1994a, hereafter RPa; RPb). In this model energy-density perturbations are generated by ground-state zero-point fluctuations during inflation (Lyth & Stewart 1990; Ratra 1994; RPa).†

Kamionkowski et al. (1994; hereafter KRSS) previously considered the open model discussed here and computed the cosmic microwave background (hereafter CMB) radiation temperature anisotropy from the generalized Sachs-Wolfe relation between the radiation perturbations and the energy-density inhomogeneity (Gouda, Sugiyama, & Sasaki 1991; Sugiyama & Silk 1994). However, KRSS normalised the model to the two-year DMR value for the rms temperature anisotropy at $10^\circ$ angular resolution (Bennett et al. 1994, hereafter B94). This KRSS normalisation technique is unsatisfactory since it fails to account for the correct DMR beam filter function and the galaxy cut as applied to the data (Wright et al. 1994a) and consequently underestimates the correct normalisation amplitude. The low observed quadrupole on the sky also affects the normalisation inferred from the rms temperature anisotropy (Banday et al. 1994).

The appropriate normalisation is determined by using all the angular information present in the data, as best represented by the power spectrum, the amplitude of which can be characterized by the rms quadrupole anisotropy amplitude, $Q_{\mathrm{rms-PS}}$. Here we use the incomplete-sky power spectrum estimation technique (described in Górski 1994, hereafter G94; and applied to the DMR data in the context of a power law CMB anisotropy model in Górski et al. 1994a, hereafter G94a) to determine $Q_{\mathrm{rms-PS}}$, as a function of $\Omega_0$, from the two-year DMR galactic sky maps. With the normalisation fixed, we then proceed to estimate various statistics of cosmological interest and subsequently comment on the viability of the open model as a description of our universe.

† In a related model where an open-inflation bubble is created by tunnelling in an inflating, spatially-flat, de Sitter spacetime, it might prove more appropriate to use a combination of the ground and excited states as an initial condition in the second inflation epoch (Bucher, Goldhaber, & Turok 1994); the final spectrum is probably determined by the model assumed for the first epoch.

## 2. SUMMARY OF COMPUTATION

The determination of the CMB power spectrum from the DMR data is hindered by the necessity to eliminate data near and on the galactic plane (because of the strong, non-cosmological, emission) from the sky maps. The proper Fourier analysis of temperature anisotropy on the cut-sky data-set should utilize orthogonal basis functions. Following the anisotropy power spectrum technique in G94, a set of orthonormal functions is constructed for the 4016 pixels surviving the $|b| > 20°$ galactic cut, and the sky temperature anisotropy is decomposed into 961 Fourier coefficients (probing the spectral range $l \leq 30$). As in G94a, both systematic errors (which are strongly limited, B94) and faint residual high-latitude galactic signals are neglected, and the combined 53 and 90 GHz two-year galactic sky maps are used in the analysis. The DMR data together with the open model multipole coefficients, parameterized by $Q_{\rm rms-PS}$ and $\Omega_0$, define an exact gaussian likelihood function $L(Q_{\rm rms-PS}, \Omega_0)$. We have computed the CMB anisotropy multipole moments, $C_\ell = \langle |a_\ell^m|^2 \rangle$, where $\delta T/T = \sum_{\ell m} a_\ell^m Y_\ell^m$, for $\Omega_0 = 0.03, 0.06$, and in steps of 0.1 from $\Omega_0 = 0.1$ to 1.0. The spectra used in this work are shown in Fig. 1, where we have adopted $\Omega_B = 0.03$ and $h = 0.5$ (the Hubble parameter $H_0 = 100h$ km s$^{-1}$ Mpc$^{-1}$). The shape of the low-$\ell$ $C_\ell$'s is almost entirely governed by $\Omega_0$ and is relatively insensitive to $\Omega_B$, $h$, and ionization history (using the tabulated values of $h$ and $\Omega_B = 0.0125h^{-2}$, for $\ell \leq 20$ we find $|C_\ell/C_2|$ is shifted by $< 3\%$ for $0.1 \leq \Omega_0 \leq 0.5$ and $\Omega_0 = 1$, and leaves the inferred normalisation essentially unchanged). The monopole and dipole are algebraically excluded from the analysis (as described in G94). We have analyzed the data both including and excluding the observed quadrupole moments, since these are the harmonics most likely to be contaminated by residual galactic and systematic emission signals. The results of this likelihood analysis are summarized in Figs. 2 – 4 and Table 1 (columns 3 and 4). Fig. 2 shows the likelihood surface for the more restrictive case with the $\ell = 2$ mode included in the analysis (with the usual assumption of uninformative, uniform prior in $Q_{rms-PS}$ and $\Omega_0$). When the quadrupole is excluded the likelihood amplitude shows less suppression near $\Omega_0 \sim 0.15$. In both cases the 95% contours of the likelihood extend throughout the entire range of $\Omega_0$ and hence no models can be formally rejected.

Using the normalisation derived from the DMR data, we determine the rms values for a number of measures of cosmological structure (see Table 1). For each value of $\Omega_0$ we record in column (2) a value for $h$ which results in an expansion time $t_0 \sim 12$Gyr (the last line is the Einstein-de Sitter case with $t_0 \sim 13$Gyr), consistent with the lower end of recent determinations (Renzini 1993; Tonry 1993), and take $\Omega_B = 0.0125h^{-2}$, the standard nucleosynthesis estimate (e.g., Reeves 1994). The quoted ranges for the tabulated statistics only account for the uncertainty in the inferred model normalisation, but ignore the uncertainty in the values of $h$ and $\Omega_B$ as derived above. Such uncertainties do not significantly affect the $Q_{\rm rms-PS}$ inference, but do affect estimates of the other statistics. Future work should more rigorously account for the allowed parameter range, as the accuracy of the $Q_{\rm rms-PS}$ determination improves. For the time being, we normalise the model to $Q_{\rm rms-PS}$ † from the no-quadrupole case (Table 1, column 4).

Column (5) gives the present rms linear mass fluctuation, $(\delta M/M)$, within an $8h^{-1}$ Mpc sphere, determined from a numerical integration of the perturbation equations, where the range here [and in columns (6) – (11)] only accounts for the $1\sigma$ uncertainty in the estimate of $Q_{\rm rms-PS}$. Columns (6), (8), and (10) give the present rms value $v_p^{(r)}$ of $\sqrt{3}$ (from isotropy) times the (gauge-invariant) radial peculiar velocity (RPa, eq. [11.59]) averaged over spheres of radius $40h^{-1}$, $60h^{-1}$ and $100h^{-1}$ Mpc, computed from a numerical integration of the perturbation equations. However, the observed large-scale velocities (which *are* radial) are conventionally quantified in terms of the rms value $v_p$ of the three-dimensional peculiar velocity. Since we cannot express $v_p$ as a sum over modes in an open space, we use the flat space expression (Peebles 1993, hereafter PPC, eq. [21.68]), with the appropriate open space power spectrum and window function, and give $v_p$ averaged over spheres of radius $40h^{-1}$, $60h^{-1}$, and $100h^{-1}$ Mpc in columns (7), (9), and (11). Interestingly, the computed values of $v_p^{(r)}$ and $v_p$ are somewhat discrepant, contrary to intuitive expectation on the scales considered here (much smaller than the curvature radius). We believe that this phenomenon merits further investigation.

---

† These numerical values do not significantly alter the reheating redshift estimated in RPb with a different assumed $Q_{\rm rms-PS}$ normalisation; we note that unlike some inflation models, in this model the observations do not impose a stringent upper bound on the slope of the inflation epoch scalar field potential (Ratra 1992; RPa).

## 3. DISCUSSION

The likelihoods shown in Fig. 3 indicate that the DMR data alone cannot rule out any value of $\Omega_0$ (including the quadrupole, $\Omega_0 \sim 0.15$ is acceptable at $> 5\%$, Fig. 2, and this is the most restrictive case). While the DMR data is consistent with the low-$\ell$ $C_\ell$'s in the scale-invariant Einstein-de Sitter model (G94a, Wright et al. 1994b), it is also consistent with a flat $\Lambda$-dominated case (Bunn & Sugiyama 1994; Górski, Stompor, & Banday 1995), and, as shown here, with an open model. Thus, the two-year DMR data do not necessarily require a flat universe or a scale-invariant model. The diminutive (galaxy-subtracted) CMB quadrupole (B94) might be considered evidence against the flat, scale-invariant model, and weakly supportive of the open models with suppressed quadrupole (Figs. 1 – 3). Nevertheless, it is as yet premature to draw such conclusions, based solely on the DMR data, since it is precisely in the very low order multipoles that the cosmic variance in all models is large and the galactic foreground contribution remains uncertain.

The rms fluctuation in the number of galaxies in a randomly placed sphere of radius $8h^{-1}$Mpc, $(\delta N/N) = 0.79$ to 1.1 (PPC, eqs. [7.33, 7.73]). From column (5), with $(\delta N/N) = b(\delta M/M)$, where $b$ is the linear bias factor, $\Omega_0 \lesssim 0.2$ requires an unreasonably high bias, $\Omega_0 \sim 0.3 - 0.5$ could be consistent with mild bias, while $\Omega_0 \gtrsim 0.7$ needs antibiasing. For $\Omega_0 \sim 0.3 - 0.5$, $(\delta M/M)[8h^{-1}\mathrm{Mpc}]$ more consistently reproduces the observed cluster mass and correlation functions (e.g., Bahcall & Cen 1992; White, Efstathiou, & Frenk 1993; although $\Omega_0 = 0.3$ might require a somewhat higher $h$ and lower $\Omega_B$ than used here). Furthermore, the shape of the observed galaxy power spectrum is better fitted by a low-density CDM model. The CDM power spectrum shape parameter corrected for $\Omega_B$ dependence, $\Omega_0 h e^{-2\Omega_B}$ (e.g., Efstathiou, Sutherland, & Maddox 1990; Peacock & Dodds 1994), is determined from miscellaneous galaxy and galaxy cluster catalogs to be $\sim 0.255 \pm 0.017$. Interestingly, this is consistent with the models in Table 1 for $\Omega_0 \sim 0.3 - 0.4$ and $h \sim 0.65$ (which is in the range now under discussion — see Freedman et al. 1994, and references therein).

Observed large scale flows, as currently described in the literature, are problematic for the open model. The POTENT analysis of Bertschinger et al. (1990) has rendered $254 < v_p(40h^{-1}\mathrm{Mpc})/(\mathrm{km\,s^{-1}}) < 522$, $163 < v_p(60h^{-1}\mathrm{Mpc})/(\mathrm{km\,s^{-1}}) < 491$, (95% c.l.). Accounting for the cosmic variance in the measurement of one bulk velocity (Bond 1994, eqs. [47, 48]) suggests that one need not yet reject any models at $\Omega_0 \gtrsim 0.3$. On scales $\sim 100h^{-1}$ Mpc the corresponding range is $333 < v_p/(\mathrm{km\,s^{-1}}) < 1045$ (Lauer & Postman 1994), implying $\Omega_0 \gtrsim 0.7$. Hence, the observed velocity fields can significantly constrain low-density models, however, a more rigorous analysis (that accounts for the survey window function and the hypersurface dependence of the velocity) as well as more and better data is required to draw a firm conclusion.

Dynamical estimates of mass clustered on scales $\lesssim 10h^{-1}$Mpc indicate $0.1 \lesssim \Omega_0 \lesssim 0.3$ (PPC, §20). The situation on scales $\gtrsim 20h^{-1}$Mpc is not yet entirely clear: if galaxies trace mass, the IRAS/POTENT analysis indicates $\Omega_0 > 0.5$ at the 95% confidence level (Dekel 1994), but some other analyses are also consistent with lower $\Omega_0$ (e.g., Roth 1993; Dekel 1994). Nusser & Davis (1994) have recently presented an estimation of $\beta_{\rm I}$ ($\equiv {\Omega_0}^{0.6}/b_{\rm IRAS} = 1.3{\Omega_0}^{0.6}(\delta M/M)[8h^{-1}{\rm Mpc}]$, where $b_{\rm IRAS}$ is the linear bias factor for IRAS galaxies) and find, at $\sim$95% confidence, $0.2 < \beta_{\rm I} < 1.0$. This, interestingly enough, corresponds to models with $0.3 \lesssim \Omega_0 \lesssim 0.6$ in Table 1.

The DMR detection of CMB anisotropy at a level significantly higher than expected in the biased scale-invariant flat CDM model focussed attention on alternative $\Omega_0 = 1$ models, such as 'tilted' CDM or mixed dark matter (MDM) models, which appear consistent with present epoch observations. However, when normalised to the DMR data, these models tend to form galaxies too recently. In the tilted case Cen & Ostriker (1993, hereafter CO) find that when $(\delta M/M)[8h^{-1}{\rm Mpc}] = 0.5$ galaxy formation peaks at a redshift $z_g \sim 0.5$ (since CO used a somewhat large $(\delta M/M)$, the upward revision of $Q_{\rm rms-PS}$ will not raise $z_g$ significantly), contrary to observational evidence suggesting that the giant elliptical luminosity functions at $z = 1$ and 0 are not dissimilar (e.g., Cowie, Songaila, & Hu 1993). The MDM model appears to predict too few damped Lyman-$\alpha$ systems (e.g., Ma & Bertschinger 1994), which are thought to be young galaxies, and are seen out to redshifts as high as $z \sim 3$ (Lanzetta 1993). It seems encouraging that adiabatic low-density models, such as the $\Lambda$-dominated flat model and the open model considered here, as well as low-density isocurvature models (Peebles 1994) and higher-bias low-density topological defect models (Pen & Spergel 1994) tend to form structure early. For instance, scaling from the CO result for the tilted case with $z_g = 0.5$, the $\Omega_0 = 0.4$ open model with $(\delta M/M)[8h^{-1}{\rm Mpc}] = 0.8$ would have $z_g = 2.5$ (e.g., RPb), which may be adequate if objects like the 'young cluster' at $z = 1$ (Le Fèvre et al. 1994) turn out to be common.

In summary, we have used the DMR two-year galactic maps to normalise an open inflation model, and have computed large-scale structure statistics in this model. The DMR data alone do not strongly constrain $\Omega_0$, but somewhat disfavour $\Omega_0 \sim 0.15$ (which is consistent with some non-CMB observations). For $\Omega_0 \sim 0.3 - 0.4$ the open model appears to be mostly consistent with observations, albeit not too impressively so. More detailed modelling (particularly of large scale flows and small angle CMB anisotropies) and improved observational constraints will be required to resolve definitively the compatibility of this model with the observed properties of the universe.

We acknowledge the efforts of those contributing to the COBE DMR. We thank D. Bond and M. Davis for helpful discussions. This work was supported in part by the Office of Space Sciences of NASA Headquarters, by NSF grant PHY89-21378, by the David and Lucile Packard Foundation, and by a JSPS Postdoctoral Fellowship for Research Abroad.

## FIGURE CAPTIONS

Fig. 1.– CMB anisotropy multipole coefficients for the open inflation model with $h = 0.5$, $\Omega_B = 0.03$, and $\Omega_0 = 0.03$, 0.05, 0.1, 0.2, 0.3, 0.4, 0.5, 0.6, 0.7, 0.8, 0.9, and 1 in ascending order, determined numerically from a Boltzmann transfer code. The spectra are normalised relative to the $C_2$ quadrupole amplitude and offset from each other to aid visualisation. At low $\ell$ the shape of the $C_\ell$ spectrum is determined by $\Omega_0$ (which fixes the perturbation growth rate and the ratio of the Hubble and space curvature scales), and by the wavelength dependence of the energy-density power spectrum.

Fig. 2.– The likelihood function $L(Q_{\rm rms-PS}, \Omega_0)$ (normalised, for clarity, to 0.9 at the highest peak near $\Omega_0 = 0.03$, $Q_{\rm rms-PS} \sim 15\mu$K) derived from a simultaneous analysis of the DMR 53 and 90 GHz two-year data for Fourier modes $\ell \in [2, 30]$. Excluding the quadrupole results in a qualitatively similar result at large and very small $\Omega_0$, with the likelihood less suppressed at $\Omega_0 \sim 0.15$. The contour plot of the likelihood shows the 99.9% (outermost), 95%, and 68% (innermost, disconnected) contour lines, and the projected location of the highest likelihood peak. One should note, on comparing to Fig. 1, that models with suppressed low-$\ell$ multipoles have somewhat higher likelihood (at low $\Omega_0$ and $\Omega_0 \sim 0.7$) than those with enhanced low-$\ell$ power ($\Omega_0 \sim 0.2$). However, the relative differences in the amplitude of the likelihood are too small to allow definite rejection of any value of $\Omega_0$.

Fig. 3.– Likelihood densities for $\Omega_0$, derived from $L(Q_{\rm rms-PS}, \Omega_0)$ (and with arbitrary equal-integrated-area normalisation). Light type shows the projected likelihood and heavy type the marginal likelihood $\left[\propto \int dQ_{\rm rms-PS} L(Q_{\rm rms-PS}, \Omega_0)\right]$. The solid lines are the case including the quadrupole, the dotted lines are the quadrupole excluded case.

Fig. 4.– Conditional likelihood densities for $Q_{\rm rms-PS}$, derived from $L(Q_{\rm rms-PS}, \Omega_0)$ (and with arbitrary equal-integrated-area normalisation), for $\Omega_0 = 1$, 0.5, 0.4, 0.3, 0.2 running from left to right. Solid lines are the case with the quadrupole, dotted lines are the no-quadrupole case.

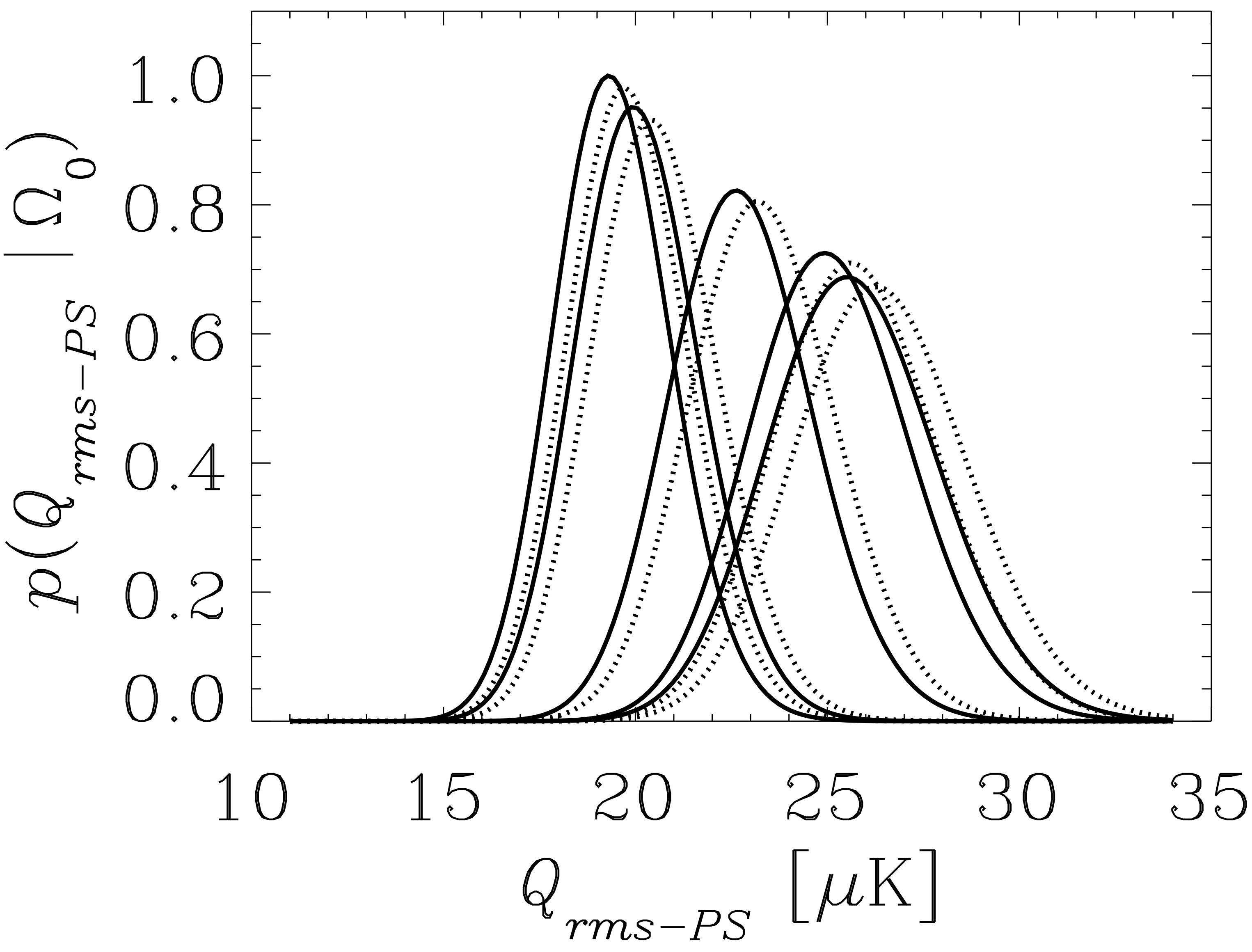

TABLE 1

Numerical Values

| $\Omega_0$ | $h$ | $Q_{\rm rms-PS}^{(Q)}$ ($\mu$K) | $Q_{\rm rms-PS}^{({\rm no}-Q)}$ ($\mu$K) | $\frac{\delta M}{M}\|_{8h^{-1}{\rm Mpc}}$ | $v_p^{(r)}\|_{40h^{-1}{\rm Mpc}}$ (km s$^{-1}$) | $v_p\|_{40h^{-1}{\rm Mpc}}$ (km s$^{-1}$) | $v_p^{(r)}\|_{60h^{-1}{\rm Mpc}}$ (km s$^{-1}$) | $v_p\|_{60h^{-1}{\rm Mpc}}$ (km s$^{-1}$) | $v_p^{(r)}\|_{100h^{-1}{\rm Mpc}}$ (km s$^{-1}$) | $v_p\|_{100h^{-1}{\rm Mpc}}$ (km s$^{-1}$) |
|---|---|---|---|---|---|---|---|---|---|---|
| 0.1 | 0.75 | $22.5 \pm 1.9$ | $23.1 \pm 2.0$ | 0.091 – 0.11 | 16 – 19 | 35 – 42 | 17 – 20 | 32 – 38 | 18 – 21 | 28 – 33 |
| 0.2 | 0.7 | $25.7 \pm 2.2$ | $26.5 \pm 2.3$ | 0.27 – 0.32 | 59 – 70 | 96 – 110 | 59 – 70 | 83 – 99 | 55 – 65 | 66 – 78 |
| 0.3 | 0.65 | $25.1 \pm 2.1$ | $25.8 \pm 2.1$ | 0.45 – 0.54 | 110 – 130 | 160 – 190 | 110 – 130 | 140 – 160 | 95 – 110 | 100 – 120 |
| 0.4 | 0.65 | $22.8 \pm 1.9$ | $23.4 \pm 1.9$ | 0.69 – 0.81 | 180 – 210 | 230 – 270 | 170 – 200 | 190 – 220 | 140 – 160 | 140 – 160 |
| 0.5 | 0.6 | $20.1 \pm 1.6$ | $20.5 \pm 1.6$ | 0.83 – 0.98 | 240 – 280 | 290 – 340 | 210 – 250 | 230 – 270 | 170 – 200 | 170 – 200 |
| 0.6 | 0.6 | $17.8 \pm 1.4$ | $18.1 \pm 1.4$ | 1.0 – 1.2 | 300 – 350 | 350 – 410 | 260 – 310 | 280 – 320 | 210 – 240 | 190 – 230 |
| 0.7 | 0.6 | $16.7 \pm 1.3$ | $17.0 \pm 1.3$ | 1.2 – 1.4 | 360 – 420 | 400 – 470 | 310 – 360 | 310 – 360 | 240 – 280 | 210 – 250 |
| 0.8 | 0.55 | $17.0 \pm 1.3$ | $17.3 \pm 1.3$ | 1.2 – 1.4 | 390 – 450 | 430 – 500 | 330 – 390 | 330 – 380 | 250 – 300 | 230 – 260 |
| 0.9 | 0.55 | $18.1 \pm 1.4$ | $18.5 \pm 1.4$ | 1.3 – 1.6 | 430 – 500 | 460 – 530 | 360 – 420 | 350 – 410 | 270 – 320 | 240 – 280 |
| 1 | 0.55 | $19.4 \pm 1.5$ | $19.8 \pm 1.6$ | 1.4 – 1.6 | 460 – 540 | 470 – 550 | 380 – 440 | 360 – 420 | 280 – 320 | 240 – 280 |
| 1 | 0.5 | $19.4 \pm 1.5$ | $19.8 \pm 1.6$ | 1.3 – 1.5 | 430 – 500 | 450 – 530 | 360 – 420 | 350 – 410 | 270 – 310 | 230 – 270 |
| (1) | (2) | (3) | (4) | (5) | (6) | (7) | (8) | (9) | (10) | (11) |